\documentclass[aps,prd,preprint,groupedaddress]{revtex4}

\usepackage{graphicx}
\usepackage{axodraw}

\def\dfrac#1#2{\displaystyle\frac{#1}{#2}}

\newcommand{\kslash}{k\kern-1ex /}
\newcommand{\pslash}{p\kern-1ex /}
\newcommand{\qslash}{q\kern-1ex /}
\newcommand{\lslash}{l\kern-1ex /}
\newcommand{\sslash}{s\kern-1ex /}
\newcommand{\Dslash}{{\cal D}\kern-1.5ex /}
\newcommand{\bpsi}{\overline{\psi}}

\newcommand{\tr}{{\rm tr}}

\newcommand{\beqa}{\begin{eqnarray}}
\newcommand{\eeqa}{\end{eqnarray}}

\newcommand{\be}{\begin{equation}}
\newcommand{\ee}{\end{equation}}
\newcommand{\ben}{\begin{eqnarray}}
\newcommand{\een}{\end{eqnarray}}
\newcommand{\nn}{\nonumber}
\def\lsim{\raise0.3ex\hbox{$<$\kern-0.75em\raise-1.1ex\hbox{$\sim$}}}
\def\gsim{\raise0.3ex\hbox{$>$\kern-0.75em\raise-1.1ex\hbox{$\sim$}}}
\def\simgt{\rlap{\lower 3.5 pt\hbox{$\mathchar \sim$}}\raise 1pt \hbox {$>$}}
\def\simlt{\rlap{\lower 3.5 pt\hbox{$\mathchar \sim$}}\raise 1pt \hbox {$<$}}

\newcommand{\mf}{{\rm MF}}
\newcommand{\tad}{{\rm tad}}
\newcommand{\msbar}{{\overline {\rm MS}}}

\newcommand{\csw}{{c_{\rm SW}}}
\newcommand{\hk}{{\hat k}}
\newcommand{\intlat}{{\int_{-\pi}^{\pi}\frac{d^4 k}{(2\pi)^4}}}

\begin{document}


\title{Determination of the Improvement Coefficient $\csw$\\
up to One-Loop Order \\with the Conventional Perturbation Theory}


\author{Sinya Aoki$^a$ and Yoshinobu Kuramashi$^b$}
\affiliation{$^a$Institute of Physics, University of Tsukuba, 
Tsukuba, Ibaraki 305-8571, Japan \\
$^b$Institute of Particle and Nuclear Studies,
High Energy Accelerator Research Organization(KEK),
Tsukuba, Ibaraki 305-0801, Japan}


\date{\today}

\begin{abstract}
We calculate the $O(a)$ improvement coefficient $\csw$ in the
Sheikholeslami-Wohlert quark action for various improved gauge actions
with six-link loops.
We employ a conventional perturbation theory introducing the 
fictitious gluon mass to regularize the infrared divergence.
Our results for some improved gauge actions are in agreement with those
previously obtained with the Schr{\"o}dinger functional method. 
\end{abstract}


\maketitle


\section{Introduction}
\label{sec:intro}

Recently the CP-PACS collaboration 
shows by a large scale of simulation 
that the hadron spectra in the quenched approximation 
systematically deviate from the experimentally observed ones
both in the meson and the baryon sectors\cite{quench}. 
It is now obvious that the next step is to incorporate the effects
of dynamical quarks to reproduce the correct hadron spectra.
With the current computational resources, however, 
unquenched QCD 
simulations are often restricted on lattices with the
lattice spacing coarser than 0.1 fm while keeping the physical volume 
larger than 2 fm.
 
A practical way to reduce the scaling violation effects 
is to employ the improved quark and gauge actions.
For the quark part the $O(a)$ improved action proposed by
Sheikholeslami and Wohlert\cite{sw} 
is now widely used. This action requires only one new term
called a clover term. Although from a theoretical point of view  
the plaquette gauge action  is already $O(a)$ improved,
a comparative numerical study employing the various quark and gauge actions 
shows that the renormalization group (RG) improved
gauge action reduces non-negligible $O(a^2)$ errors\cite{compara}.
Moreover, JLQCD collaboration has recently reported 
that the first order phase transition
observed in the three flavor QCD simulation 
with the $O(a)$ improved quark action and the plaquette gauge action,
which is considered to be a lattice artifact, 
disappears once the
gauge action is replaced by the RG improved one\cite{nf3}. 
Thus the improvement of the gauge action is mandatory for
the three flavor QCD simulation at the currently accessible lattice spacing. 

In this paper we determine
the clover coefficient $\csw$ in the massless
SW quark action up to one-loop order for various improved gauge actions
including the DBW2 action\cite{dbw2}.
Preparing for new improved gauge actions yet to come,  
we parameterize the value of $\csw$ as a function of
the improvement coefficient of gauge action
for later convenience. 
Another important purpose of the present calculation is to check
the validity of the conventional perturbative method 
for the determination of the massless clover coefficient $\csw$. 
Although previous calculations of $\csw$ are 
done by the twisted antiperiodic boundary conditions\cite{csw_w} or 
the Schr{\"o}dinger functional method\cite{csw_sf}, 
we instead employ the conventional perturbation theory with the use of the 
fictitious gluon mass to regularize the infrared divergence, 
which has been applied successfully for the calculation 
of the renormalization constants and the
improvement coefficients for the bilinear quark operators\cite{gmass}.
This method can be easily implemented, within the standard knowledge of
perturbation theory.
Our results for some improved gauge actions are in agreement with those
previously obtained with the Schr{\"o}dinger functional method,
which assures the validity of our conventional perturbative method.
We are now extending this calculation of $\csw$ to the case of
the heavy quark formulation proposed by the authors\cite{akt}, 
where the conventional perturbative method 
is much easier to handle the massive quarks
than the Schr{\"o}dinger functional method. 

This paper is organized as follows. In Sec.~II we introduce the 
improved quark and gauge actions and their Feynman rules relevant for the
present calculation. In Sec.~III we determine the clover coefficient 
$\csw$ up to one-loop level 
from the on-shell quark-quark scattering amplitude.
The result of $\csw$ is parametrized as a function of
the improvement coefficient of the gauge action. 
Our conclusions are summarized in Sec.~IV. 

The physical quantities are expressed in lattice units and 
the lattice spacing $a$ is suppressed unless necessary.
We take  SU($N_c$) gauge group with the gauge coupling constant $g$.

\section{Action and Feynman Rules}
\label{sec:action}
 
For the quark action we consider the $O(a)$-improved quark action\cite{sw}:
\ben
S_{\rm quark}&=&
\sum_n\frac{1}{2}\sum_\mu
\left\{{\bar \psi}_n(-r+\gamma_\mu)U_{n,\mu}\psi_{n+{\hat \mu}}
      +{\bar \psi}_n(-r-\gamma_\mu)U^\dagger_{n-{\hat \mu},\mu}
       \psi_{n-{\hat \mu}}\right\}
+(m_0+4r)\sum_n{\bar \psi}_n\psi_n \nn\\
&&-\csw\sum_n\sum_{\mu,\nu}ig\frac{r}{4}
{\bar \psi}_n\sigma_{\mu\nu}F_{\mu\nu}(n)\psi_n,
\label{eq:action_q}
\een
where we define the Euclidean gamma matrices in terms of
the Minkowski matrices in the Bjorken-Drell convention:
$\gamma_j=-i\gamma_{BD}^j$ $(j=1,2,3)$, 
$\gamma_4=\gamma_{BD}^0$,
$\gamma_5=\gamma_{BD}^5$ and 
$\sigma_{\mu\nu}=\frac{1}{2}[\gamma_\mu,\gamma_\nu]$.
The field strength $F_{\mu\nu}$ in the clover term
is given by
\ben 
F_{\mu\nu}(n)&=&\frac{1}{4}\sum_{i=1}^{4}\frac{1}{2ig}
\left(U_i(n)-U_i^\dagger(n)\right), \\
U_1(n)&=&U_{n,\mu}U_{n+{\hat \mu},\nu}
         U^\dagger_{n+{\hat \nu},\mu}U^\dagger_{n,\nu}, \\
U_2(n)&=&U_{n,\nu}U^\dagger_{n-{\hat \mu}+{\hat \nu},\mu}
         U^\dagger_{n-{\hat \mu},\nu}U_{n-{\hat \mu},\mu}, \\
U_3(n)&=&U^\dagger_{n-{\hat \mu},\mu}U^\dagger_{n-{\hat \mu}-{\hat \nu},\nu}
         U_{n-{\hat \mu}-{\hat \nu},\mu}U_{n-{\hat \nu},\nu}, \\
U_4(n)&=&U^\dagger_{n-{\hat \nu},\nu}U_{n-{\hat \nu},\mu}
         U_{n+{\hat \mu}-{\hat \nu},\nu}U^\dagger_{n,\mu}.
\een
The weak coupling perturbation theory is developed  
by writing the link variable in terms of the gauge potential
\be
U_{n,\mu}=\exp \left(igaT^A A_\mu^A\left(n+\frac{1}{2}\hat\mu\right)\right),
\ee
where $T^A$ ($A=1,\dots,N^2_c-1$) is a generator of color SU($N_c$).

The quark propagator is obtained by inverting Wilson Dirac operator 
in eq.(\ref{eq:action_q}),
\be
S_q^{-1}(p)=i\sum_\mu \gamma_\mu {\rm sin}(p_\mu)+m_0+
r\sum_\mu(1-{\rm cos}(p_\mu)).
\ee
To calculate the improvement coefficient $\csw$ 
up to one-loop level, we need 
one-, two- and three-gluon vertices with quarks:
\ben
V^A_{1\mu}(p,q)&=&-gT^A\left\{
i\gamma_\mu{\rm cos}\left(\frac{p_\mu+q_\mu}{2}\right)
+r{\rm sin}\left(\frac{p_\mu+q_\mu}{2}\right)\right\}, 
\label{eq:vertex_w1}\\
V_{2\mu\nu}^{AB} (p,q)
&=& \frac{a}{2} g^2 \frac{1}{2} \{T^{A}, T^{B}\}\delta_{\mu\nu}
\left\{ i \gamma_\mu \sin \left(\frac{p_\mu+q_\mu}{2}\right)
-r \cos \left(\frac{p_\mu+q_\mu}{2}\right)\right\},
\label{eq:vertex_w2}\\
V_{3\mu\nu\tau}^{ABC} (p,q)
&=& \frac{a^2}{6} g^3 \frac{1}{6} 
\left[T^{A}\{T^{B},T^{C}\}+T^{B}\{T^{C},T^{A}\}+T^{C}\{T^{A},T^{B}\}\right]
\delta_{\mu\nu}\delta_{\mu\tau}\nn\\
&&\times\left\{i\gamma_\mu \cos \left(\frac{p_\mu+q_\mu}{2}\right)
+r \sin \left(\frac{p_\mu+q_\mu}{2}\right)\right\},
\label{eq:vertex_w3}\\
V^A_{c1\mu}(p,q)&=&-gT^A\csw\frac{r}{2}
\sum_\nu \sigma_{\mu\nu}
\cos \left(\frac{p_\mu-q_\mu}{2}\right)\sin (p_\nu-q_\nu),
\label{eq:vertex_c1}\\
V^{AB}_{c2\mu\nu}(p,q,k_1,k_2)&=&
-\frac{a}{2} g^2 if_{ABC}T^C\csw\frac{r}{4}\nn\\
&&\times\left\{
\sigma_{\mu\nu}\left[
4\cos \left(\frac{k_{1\nu}}{2}\right)
\cos \left(\frac{k_{2\mu}}{2}\right)
\cos \left(\frac{q_\mu-p_\mu}{2}\right)
\cos \left(\frac{q_\nu-p_\nu}{2}\right)\right.\right.\nn\\
&&\left.\left.
-2\cos \left(\frac{k_{1\mu}}{2}\right)
\cos \left(\frac{k_{2\nu}}{2}\right)\right]\right.\\
&&\left.+\delta_{\mu\nu}\sum_\rho\sigma_{\mu\rho}
\sin \left(\frac{q_\mu-p_\mu}{2}\right)
\left[\sin(k_{2\rho})-\sin(k_{1\rho})\right]\right\},\nn
\label{eq:vertex_c2}\\
V^{ABC}_{c3\mu\nu\tau}(p,q,k_1,k_2,k_3)
&=&-3ig^3 \frac{a^2}{6}\csw r\nn\\
&&\times\left[ T^{A}T^{B}T^{C}
\delta_{\mu\nu}\delta_{\mu\tau}
\sum_\rho i\sigma_{\mu\rho}\left\{ 
-\frac{1}{6}\cos \left(\frac{q_\mu-p_\mu}{2}\right)
\sin (q_\rho-p_\rho)\right.\right.\nn\\
&&\left.\left.
+\cos \left(\frac{q_\mu-p_\mu}{2}\right)
\cos \left(\frac{q_\rho-p_\rho}{2}\right)
\cos \left(\frac{k_{3\rho}-k_{1\rho}}{2}\right)
\sin \left(\frac{k_{2\rho}}{2}\right)
\right\}\right.\nn\\
&&\left.
-\frac{1}{2}\left[T^A T^B T^C+T^C T^B T^A\right]i\sigma_{\mu\nu}\right.\\
&&\left.
\times\left\{\delta_{\nu\tau}
2\cos \left(\frac{q_\mu-p_\mu}{2}\right)
\cos \left(\frac{q_\nu-p_\nu}{2}\right)
\cos \left(\frac{k_{3\mu}+k_{2\mu}}{2}\right)
\sin \left(\frac{k_{1\nu}}{2}\right)\right.\right.\nn\\
&&\left.\left.
+\delta_{\nu\tau}
\sin \left(\frac{k_{3\nu}+k_{2\nu}}{2}\right)
\cos \left(\frac{k_{1\mu}}{2}+k_{2\mu}\right)\right.\right.\nn\\
&&\left.\left.
+\delta_{\mu\tau}
\sin \left(\frac{k_{1\mu}+2k_{2\mu}+k_{3\mu}}{2}\right)
\cos \left(\frac{q_\nu-p_\nu}{2}\right)
\cos \left(\frac{k_{3\nu}-k_{1\nu}}{2}\right)
\right\}\right],\nn
\label{eq:vertex_c3}
\een
where $f_{ABC}$ the structure constant of SU($N_c$) gauge group.
The first three vertices originate from 
the Wilson quark action and the last three from the clover term. 
The momentum assignments for the vertices are depicted 
in Fig.~\ref{fig:fr_vtx_qg}.

For the gauge  action we consider the following
general form including the standard plaquette term 
and six-link loop terms:
\be
S_{\rm g}=\frac{1}{g^2}\left\{ c_0\sum_{\rm plaquette}\tr U_{pl}
+c_1\sum_{\rm rectangle}\tr U_{rtg}
+c_2\sum_{\rm chair}\tr U_{chr}
+c_3\sum_{\rm parallelogram} \tr U_{plg} \right\}
\label{eq:action_g}
\ee
with the normalization condition
\be
c_0+8c_1+16c_2+8c_3=1,
\ee
where six-link loops are composed of a $1\times 2$ rectangle,
a bent $1\times 2$ rectangle (chair) and a three-dimensional
parallelogram.
In this paper we consider the following choices:
$c_1=c_2=c_3=0$(Plaquette),
$c_1=-1/12$, $c_2=c_3=0$(Symanzik)\cite{Weisz83,LW} 
$c_1=-0.331$, $c_2=c_3=0$(Iwasaki), $c_1=-0.27, 
c_2+c_3=-0.04$(Iwasaki')
\cite{Iwasaki83}, $c_1=-0.252, c_2+c_3=-0.17$(Wilson)\cite{Wilson}
and $c_1 = -1.40686$, $c_2=c_3=0$(DBW2)\cite{dbw2}.
The last four cases are called the RG improved gauge action, whose 
parameters are chosen to be the values 
suggested by approximate renormalization group analyses.
Some of these actions are now getting widely used, since
they realize continuum-like gauge field fluctuations 
better than the naive plaquette action at the same lattice spacing.

The free gluon propagator is derived in Ref.~\cite{Weisz83}:
\ben
D_{\mu\nu}(k)&=&\frac{1}{(\hk^2)^2}\left[
(1-A_{\mu\nu})\hk_\mu\hk_\nu+
\delta_{\mu\nu}\sum_\sigma\hk_\sigma^2 A_{\nu\sigma}\right]
\een
with 
\ben
\hk_\mu&=&2{\rm sin}\left(\frac{k_\mu}{2}\right),\\
\hk^2&=&\sum_{\mu=1}^{4}\hk_\mu^2,
\een
where we employ the Feynman gauge.
The matrix $A_{\mu\nu}$ satisfies
\ben
&({\rm i})& A_{\mu\mu}=0\;\;\; {\rm for}\;\;{\rm all}\;\; \mu, \\
&({\rm ii})& A_{\mu\nu}=A_{\nu\mu}, \\
&({\rm iii})& A_{\mu\nu}(k)=A_{\mu\nu}(-k). \\
&({\rm iv})& A_{\mu\nu}(0)=1\;\;\; {\rm for}\;\; \mu\ne\nu,
\een
and its expression is given by
\ben
A_{\mu\nu}(k)&=&\frac{1}{\Delta_4}
\left[(\hk^2-\hk_\nu^2)(q_{\mu\rho}q_{\mu\tau}\hk_\mu^2
+q_{\mu\rho}q_{\rho\tau}\hk_\rho^2
+q_{\mu\tau}q_{\rho\tau}\hk_\tau^2)\right. \nn\\
&&\left.+(\hk^2-\hk_\mu^2)(q_{\nu\rho}q_{\nu\tau}\hk_\nu^2
+q_{\nu\rho}q_{\rho\tau}\hk_\rho^2
+q_{\nu\tau}q_{\rho\tau}\hk_\tau^2)\right. \nn\\
&&\left.+q_{\mu\rho}q_{\nu\tau}(\hk_\mu^2+\hk_\rho^2)(\hk_\nu^2+\hk_\tau^2) 
+q_{\mu\tau}q_{\nu\rho}(\hk_\mu^2+\hk_\tau^2)(\hk_\nu^2+\hk_\rho^2)\right. \nn\\
&&\left.-q_{\mu\nu}q_{\rho\tau}(\hk_\rho^2+\hk_\tau^2)^2
-(q_{\mu\rho}q_{\nu\rho}+q_{\mu\tau}q_{\nu\tau})\hk_\rho^2\hk_\tau^2\right. \nn\\
&&\left.-q_{\mu\nu}(q_{\mu\rho}\hk_\mu^2\hk_\tau^2
           +q_{\mu\tau}\hk_\mu^2\hk_\rho^2
           +q_{\nu\rho}\hk_\nu^2\hk_\tau^2
           +q_{\nu\tau}\hk_\nu^2\hk_\rho^2)\right],
\label{eq:matrix_a}
\een
with $\mu\ne\nu\ne\rho\ne\tau$ the Lorentz indices. 
$q_{\mu\nu}$ and $\Delta_4$  are written as
\ben
q_{\mu\nu}&=&(1-\delta_{\mu\nu})
\left[1-(c_1-c_2-c_3)(\hk_\mu^2+\hk_\nu^2)-(c_2+c_3)\hk^2\right], \\
\Delta_4&=&\sum_\mu \hk_\mu^4\prod_{\nu\ne\mu}q_{\nu\mu}
+\sum_{\mu >\nu,\rho >\tau,\{\rho,\tau\}\cap\{\mu,\nu\}=\emptyset}
\hk_\mu^2\hk_\nu^2q_{\mu\nu}(q_{\mu\rho}q_{\nu\tau}
                            +q_{\mu\tau}q_{\nu\rho}).
\een
In the case of the standard plaquette action,
the matrix $A_{\mu\nu}$ is simplified as
\be
A_{\mu\nu}^{\rm plaquette}=1-\delta_{\mu\nu}.
\ee

\begin{figure}[t]
\begin{center}

\begin{picture}(200,150)(0,50)
\Gluon(50,50)(100,100){-5}{5}
\Gluon(100,100)(50,150){-5}{5}
\Vertex(100,100){2}
\Gluon(100,100)(171,100){5}{5}
\LongArrow(50,65)(64,79)
\Text(25,60)[c]{$k_1$, $\lambda$, $A$}
\LongArrow(50,135)(64,121)
\Text(25,145)[c]{$k_2$, $\rho$, $B$}
\LongArrow(171,115)(151,115)
\Text(196,115)[c]{$k_3$, $\tau$, $C$}
\end{picture}

\end{center}
\caption{Momentum assignment for the three-gluon vertex.}
\label{fig:fr_vtx_3g}
\vspace{8mm}
\end{figure}
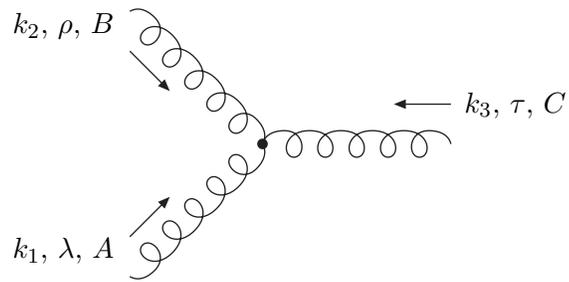

The present calculation requires only the three-gluon vertex which
is given in Ref.~\cite{Weisz83},
\be
V^{ABC}_{g3\lambda\rho\tau}(k_1,k_2,k_3)
=-i\frac{g}{6}f_{ABC}
\sum_{i=0}^{3}c_iV^{(i)}_{g3\lambda\rho\tau}(k_1,k_2,k_3)
\ee
with
\ben
V^{(0)}_{g3\lambda\rho\tau}(k_1,k_2,k_3)
&=&\delta_{\lambda\rho}\widehat{(k_1-k_2)}_\tau c_{3\lambda}
+{\rm 2\; cycl.\;\;perms.},\\
V^{(1)}_{g3\lambda\rho\tau}(k_1,k_2,k_3)
&=&8V^{(0)}_{g3\lambda\rho\tau}(k_1,k_2,k_3)\nn\\
&&+\left[ 
\delta_{\lambda\rho}\left\{ 
c_{3\lambda}(\widehat{(k_1-k_2)}_\lambda
(\delta_{\lambda\tau}\hk_3^2-\hk_{3\lambda}\hk_{3\tau})
-\widehat{(k_1-k_2)}_\tau(\hk_{1\tau}^2+\hk_{2\tau}^2))\right.
\right.
\nn\\
&+&
\left.
\left.\widehat{(k_1-k_2)}_\tau
(\hk_{1\lambda}\hk_{2\lambda}
-2c_{1\lambda}c_{2\lambda}\hk_{3\lambda}^2)\right\}+{\rm 2\; cycl.\;\;perms.}
\right]
,\\
V^{(2)}_{g3\lambda\rho\tau}(k_1,k_2,k_3)
&=&16V^{(0)}_{g3\lambda\rho\tau}(k_1,k_2,k_3)\nn\\
&&-\left[
\delta_{\lambda\rho}(1-\delta_{\lambda\tau})
c_{3\lambda}\sum_{\sigma\ne\lambda,\tau}
\left\{\widehat{(k_1-k_2)}_\tau
(\hk_{1\sigma}^2+\hk_{2\sigma}^2+\hk_{3\sigma}^2)
+\hk_{3\tau}(\hk_{1\sigma}^2-\hk_{2\sigma}^2)\right\}
\right.
\nn\\ 
&&\left.
+(1-\delta_{\lambda\rho})(1-\delta_{\lambda\tau})(1-\delta_{\rho\tau})
\hk_{1\lambda}\hk_{2\rho}\widehat{(k_1-k_2)}_\tau+{\rm 2\; cycl.\;\;perms.}
\right]
,\\
V^{(3)}_{g3\lambda\rho\tau}(k_1,k_2,k_3)
&=&8V^{(0)}_{g3\lambda\rho\tau}(k_1,k_2,k_3)\nn\\
&&-\left[
\delta_{\lambda\rho}(1-\delta_{\lambda\tau})
c_{3\lambda}\widehat{(k_1-k_2)}_\tau
\sum_{\sigma\ne\lambda,\tau}(\hk_{1\sigma}^2+\hk_{2\sigma}^2)
\right.
\nn\\
&&+\frac{1}{2}(1-\delta_{\lambda\rho})(1-\delta_{\lambda\tau})
(1-\delta_{\rho\tau})\widehat{(k_1-k_2)}_\tau
\left\{\hk_{1\lambda}\hk_{2\rho}
-\frac{1}{3}\widehat{(k_3-k_1)}_\rho\widehat{(k_2-k_3)}_\lambda \right\}\nn\\
&&\left. +{\rm 2\; cycl.\;\;perms.}
\right],
\een
where we introduce the notation,
\ben
c_{i\lambda}&=&\cos\left(\frac{k_{i\lambda}}{2}\right).
\een
The momentum assignment is found in Fig.~\ref{fig:fr_vtx_3g}.

\section{Determination of $\csw$ up to one-loop level}

\begin{figure}[t]
\begin{center}
\begin{picture}(400,100)(0,0)
\ArrowLine(150,0)(200,50)
\Text(190,20)[l]{$p$}
\ArrowLine(200,50)(150,100)
\Text(190,80)[l]{$q$}
\Vertex(200,50){2}
\Gluon(200,50)(300,50){5}{8}
\Text(250,80)[c]{$p-q$, $\mu$, $A$}
\LongArrow(240,65)(260,65)
\end{picture}
\end{center}
\caption{Quark-gluon vertex at tree level.}
\label{fig:vtx_tree}
\vspace{8mm}
\end{figure}
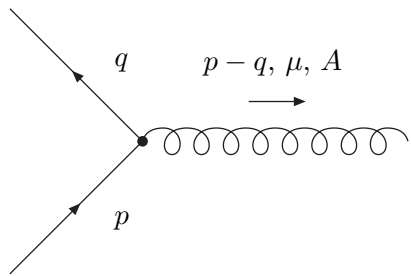

The first calculation of the clover coefficient up to the
one-loop level $\csw=\csw^{(0)}+g^2\csw^{(1)}$
was done by Wohlert\cite{csw_w}, who determined
it for the plaquette gauge action 
to eliminate the $O(a)$ contribution in the on-shell 
quark-quark scattering amplitude. 
Since the gauge propagator is already $O(a)$ improved,
the $O(a)$ contributions arise only from quark-gluon vertex.
At tree-level the quark-gluon vertex 
in Fig.~\ref{fig:vtx_tree} is written as
\ben
\Lambda_\mu^{(0)}(p,q)=-gT^A\left\{i\gamma_\mu
+r\left(\frac{p_\mu a+q_\mu a}{2}\right)\right\}
-g\frac{r\csw}{2}T^A\sum_\nu\sigma_{\mu\nu}(p_\nu-q_\nu)a
+O(a^2).
\een
where $p$ and $q$ are incoming and outgoing quark momenta assumed 
to be much less than the cutoff $a^{-1}$.
We set the Wilson parameter to $r=1$.
Sandwiching $\Lambda_\mu(p,q)$ by the Dirac spinor we obtain 
\ben
{\bar u}(q)\Lambda_\mu^{(0)}(p,q) u(p)&=&
-gT^A{\bar u}(q)
\{i\gamma_\mu+(1-\csw^{(0)})\frac{a}{2}(p_\mu+q_\mu)\}u(p)+O(a^2),
\een
where we use the Gordon identity.
We find that $\csw^{(0)}$ should be one to eliminate the $O(a)$ term.

\begin{figure}[t]
\begin{center}

\begin{picture}(450,150)(0,0)
\ArrowLine(50,50)(75,75)
\Text(70,55)[l]{$p$}
\ArrowLine(75,75)(100,100)
\Text(90,70)[l]{$p+k$}
\Vertex(75,75){2}
\ArrowLine(100,100)(75,125)
\Text(70,145)[l]{$q$}
\ArrowLine(75,125)(50,150)
\Text(90,130)[l]{$q+k$}
\Vertex(75,125){2}
\Vertex(100,100){2}
\Gluon(100,100)(200,100){5}{8}
\Text(150,130)[c]{$p-q$, $\mu$, $A$}
\LongArrow(140,115)(160,115)
\Vertex(75,75){2}
\Vertex(75,125){2}
\Gluon(75,75)(75,125){5}{4}
\LongArrow(60,110)(60,90)
\Text(50,100)[c]{$k$}
\Text(125,30)[c]{(a)}
\ArrowLine(250,50)(275,75)
\Gluon(300,100)(275,75){5}{3}
\Text(270,55)[l]{$p$}
\Text(290,70)[l]{$p-q+k$}
\Vertex(275,75){2}
\Gluon(275,125)(300,100){5}{3}
\ArrowLine(275,125)(250,150)
\Text(270,145)[l]{$q$}
\Text(290,130)[l]{$k$}
\Vertex(275,125){2}
\Vertex(300,100){2}
\Gluon(300,100)(400,100){5}{8}
\Text(350,130)[c]{$p-q$, $\mu$, $A$}
\LongArrow(340,115)(360,115)
\Vertex(275,75){2}
\Vertex(275,125){2}
\ArrowLine(275,75)(275,125)
\Text(270,100)[r]{$q-k$}
\Text(325,30)[c]{(b)}
\end{picture}

\begin{picture}(450,150)(0,0)
\ArrowLine(50,50)(100,100)
\Text(80,65)[l]{$p$}
\ArrowLine(100,100)(50,150)
\Text(80,135)[l]{$q$}
\Vertex(100,100){2}
\LongArrow(130,130)(110,130)
\Text(120,140)[c]{$k$}
\GlueArc(120,100)(20,0,180){5}{7}
\GlueArc(120,100)(20,180,360){5}{7}
\LongArrow(110,70)(130,70)
\Text(120,60)[c]{$p-q+k$}
\Vertex(140,100){2}
\Gluon(140,100)(200,100){5}{5}
\LongArrow(160,115)(180,115)
\Text(170,130)[c]{$p-q$, $\mu$, $A$}
\Text(125,30)[c]{(c)}
\ArrowLine(250,50)(300,100)
\Text(280,65)[l]{$p$}
\ArrowLine(300,100)(250,150)
\Text(280,135)[l]{$q$}
\Vertex(300,100){2}
\LongArrow(330,130)(310,130)
\Text(320,140)[c]{$k$}
\GlueArc(320,100)(20,-180,180){5}{14}
\GlueArc(350,100)(50,180,360){5}{16}
\LongArrow(380,115)(400,115)
\Text(390,130)[c]{$p-q$, $\mu$, $A$}
\Text(325,30)[c]{(d)}
\end{picture}

\begin{picture}(450,150)(0,0)
\ArrowLine(50,50)(75,75)
\Vertex(75,75){2}
\Text(45,100)[r]{$q+k$}
\ArrowArcn(75,100)(25,270,90)
\Text(70,55)[l]{$p$}
\ArrowLine(75,125)(50,150)
\Text(70,145)[l]{$q$}
\Vertex(75,125){2}
\GlueArc(75,100)(25,-90,90){5}{8}
\LongArrow(110,110)(110,90)
\Text(115,100)[l]{$k$}
\Gluon(200,75)(75,75){5}{11}
\Text(160,105)[c]{$p-q$, $\mu$, $A$}
\LongArrow(150,90)(170,90)
\Text(125,30)[c]{(e)}
\ArrowLine(250,50)(275,75)
\Vertex(275,75){2}
\ArrowArcn(275,100)(25,270,90)
\Text(270,55)[l]{$p$}
\ArrowLine(275,125)(250,150)
\Text(270,145)[l]{$q$}
\Vertex(275,125){2}
\GlueArc(275,100)(25,-90,90){5}{8}
\Text(245,100)[r]{$p+k$}
\LongArrow(310,110)(310,90)
\Text(315,100)[l]{$k$}
\Gluon(275,125)(400,125){5}{11}
\Text(360,95)[c]{$p-q$, $\mu$, $A$}
\LongArrow(350,110)(370,110)
\Text(325,30)[c]{(f)}
\end{picture}

\vspace{-8mm}
\end{center}
\caption{Quark-gluon vertex at one-loop level.}
\label{fig:vtx_1loop}
\end{figure}
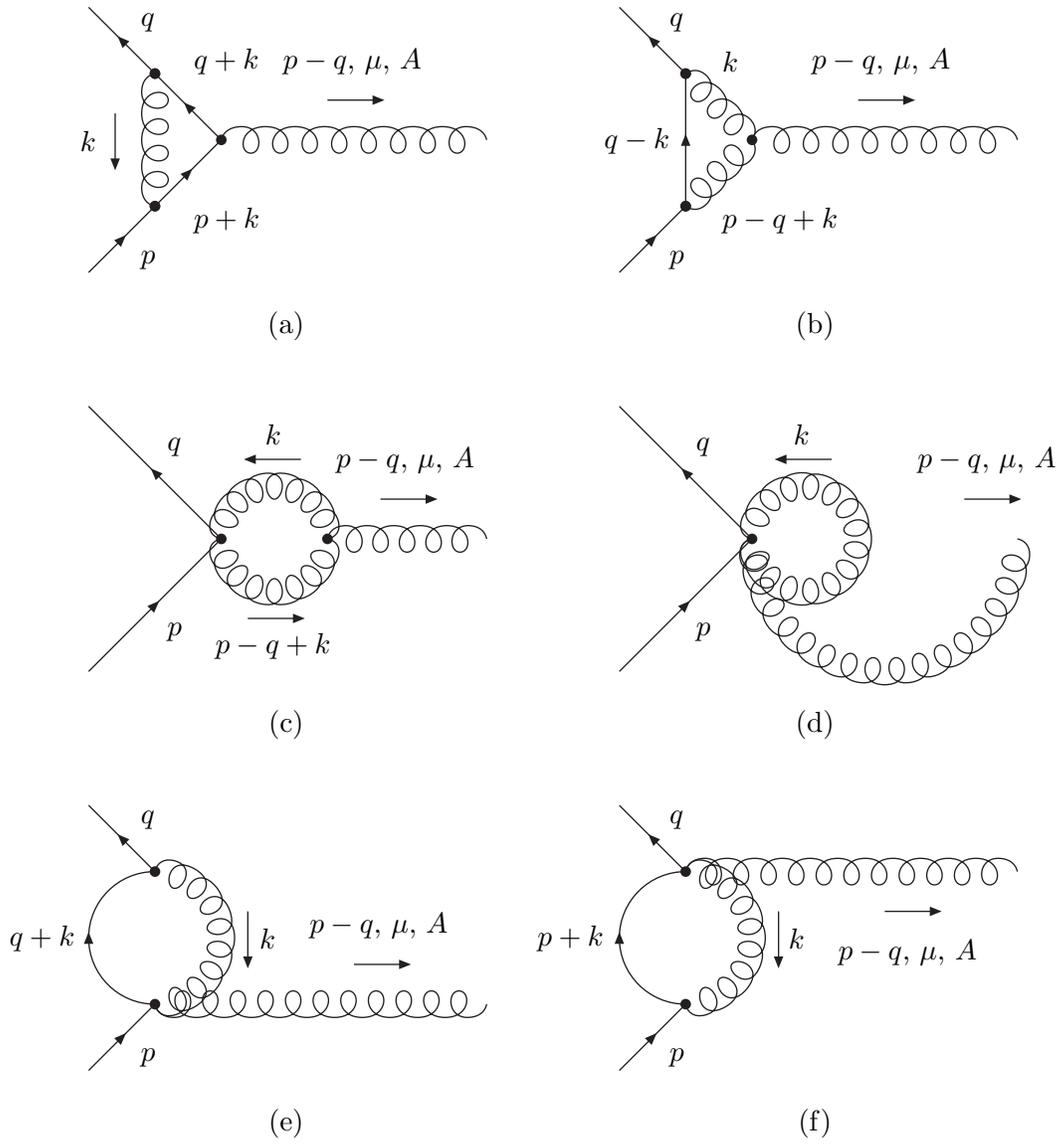

To determine the one-loop coefficient $\csw^{(1)}$ 
we need six types of diagrams shown 
in Fig.~\ref{fig:vtx_1loop}. 
The contribution of each diagram to the vertex function is denoted by
\ben
\Lambda_\mu^{(1)}(p,q) =\sum_{i=a,\dots,f}\Lambda_\mu^{(1-i)}(p,q)=
\sum_{i=a,\dots,f}\int_{-\pi}^{\pi}\frac{d^4 k}{(2\pi)^4}I_\mu^{(i)}(p,q,k). 
\een
Here we are concerned with the infrared divergences 
originating from some types of diagrams.
Although they are supposed to be canceled out after summing up the
contributions of all the diagrams, we need to introduce some
infrared regularization in the process of the calculation. 
While previous calculations employ 
the twisted antiperiodic boundary conditions\cite{csw_w} or the 
Schr{\"o}dinger functional method\cite{csw_sf} for this purpose,
we instead employ the fictitious gluon mass $\lambda$
with the ordinary perturbation theory\cite{gmass}: 
the infrared divergences 
are extracted by an analytically integrable
expression ${\tilde I}_\mu^{(i)}(p,q,k,\lambda)$ 
which has the same infrared
behavior as $I_{\mu}^{(i)}(p,q,k)$,
\ben  
\Lambda_\mu^{(1-i)}(p,q) &=& \int_{-\pi}^{\pi}\frac{d^4 k}{(2\pi)^4}
\theta(\Lambda^2-k^2)\tilde I_\mu^{(i)}(p,q,k,\lambda)\nn\\ 
&&+ \int_{-\pi}^{\pi}\frac{d^4 k}{(2\pi)^4} 
\left.\left\{I_\mu^{(i)}(p,q,k) 
-\theta(\Lambda^2-k^2)\tilde I_\mu^{(i)}(p,q,k,\lambda) 
\right\}\right|_{\lambda\rightarrow 0}
\label{eq:Lambda_subtract}
\een
with a cut-off $\Lambda$ $(\le\pi)$. The Heaviside function 
$\theta$ is introduced to restrict the domain of integration to 
a hypersphere of radius $\Lambda$, which makes the integral
analytically calculable.
Since we are interested in the $O(g^2a)$ contributions,
the counter terms ${\tilde I}_\mu^{(i)}(p,q,k,\lambda)$  
can be composed of the propagators and vertices, obtained
from an expansion of the Feynman rules in Sec.~\ref{sec:action} up to $O(a)$:
\ben
{\tilde S}_q(p)&=&\frac{-i\pslash+arp^2/2}{p^2},\\
{\tilde V}^A_{1\mu}(p,q)&=&-gT^A\left\{
i\gamma_\mu+\frac{a}{2}r(p_\mu+q_\mu)\right\}, \\
{\tilde V}_{2\mu\nu}^{AB} (p,q)
&=& \frac{a}{2} g^2 \frac{1}{2} \{T^{A}, T^{B}\}(-r)\delta_{\mu\nu},\\
{\tilde V}^A_{c1\mu}(p,q)&=&-gT^A\csw\frac{ar}{2}
\sum_\nu \sigma_{\mu\nu}(p_\nu-q_\nu),\\
{\tilde V}^{AB}_{c2\mu\nu}(p,q,k_1,k_2)&=&
-\frac{a}{2} g^2 if_{ABC}T^C\csw\frac{r}{2}
\sigma_{\mu\nu},\\
{\tilde D}_{\mu\nu}(k,\lambda)&=&\frac{\delta_{\mu\nu}}{k^2+\lambda^2},\\
{\tilde V}^{ABC}_{g3\lambda\rho\tau}(k_1,k_2,k_3)
&=&-i\frac{g}{6}f_{ABC}\left\{\delta_{\lambda\rho}(k_1-k_2)_\tau 
+{\rm 2\; cycl.\;\;perms.}\right\},
\een
where we consider the massless case. 
The momentum assignments are depicted 
in Figs.~\ref{fig:fr_vtx_qg} and \ref{fig:fr_vtx_3g}.

 From the Lorentz symmetry and the parity conservation,
the off-shell vertex function up to $O(p,q)$ is written as
\ben
\Lambda_\mu^{(1)}(p,q)&=&-g^3T^A\left\{\gamma_\mu F_1 
+a\qslash \gamma_\mu F_2+a\gamma_\mu\pslash  F_3 \right.\nn\\
&&\left.+a(p_\mu+q_\mu) G_1+a(p_\mu-q_\mu)  H_1+O(p^2,q^2,pq)\right\},
\label{eq:vtx_offshell}
\een
where $F_i$ $(i=1,2,3)$, $G_1$ and $H_1$ are dimensionless functions.
Sandwiching $\Lambda_\mu^{(1)}(p,q)$ by the on-shell quark states
the matrix elements are reduced to be
\ben
{\bar u}(q)\Lambda_\mu^{(1)}(p,q) u(p)&=&
-g^3T^A\left\{{\bar u}(q)\gamma_\mu u(p) F_1 
+a(p_\mu+q_\mu){\bar u}(q) u(p) G_1\right.\nn\\
&&\left.+a(p_\mu-q_\mu){\bar u}(q) u(p) H_1\right\}, 
\label{eq:vtx_onshell}
\een
where $\pslash u(p)=0$ and ${\bar u}(q)\qslash=0$.
 From a view point of the on-shell improvement,
the second and third terms of the right hand side 
represent the contributions of the dimension five operators,
\ben
{\cal O}_{+}&=&(\partial_\nu\bpsi(x))\sigma_{\mu\nu}\psi(x)
+\bpsi(x)\sigma_{\mu\nu}(\partial_\nu\bpsi(x)),\\
{\cal O}_{-}&=&(\partial_\nu\bpsi(x))\sigma_{\mu\nu}\psi(x)
-\bpsi(x)\sigma_{\mu\nu}(\partial_\nu\bpsi(x)).
\een
Here we should note that the transformation property of ${\cal O}_{-}$
in terms of charge conjugation is different from 
that of $\bpsi(x)\gamma_\mu\psi(x)$, which means that the 
last term of eq.(\ref{eq:vtx_onshell}) never appears, namely $H_1=0$.
 From the expression (\ref{eq:vtx_offshell}) 
we can extract the coefficient $G_1$ as 
\ben
-g^3T^AG_1&=&\left.\frac{1}{8}{\rm Tr} 
\left[\left\{\frac{\partial}{\partial p_\mu}
+\frac{\partial}{\partial q_\mu}\right\}\Lambda_\mu^{(1)}(p,q)
+\left\{\frac{\partial}{\partial p_\nu}
-\frac{\partial}{\partial q_\nu}\right\}
\Lambda_\mu^{(1)}(p,q)\gamma_\nu\gamma_\mu\right]
\right|^{\mu\ne\nu}_{p,q\rightarrow 0}
\een

It would be instructive to show how the infrared divergence 
in each diagram cancels out after the summation.
Let us take the case of the plaquette gauge action as an example. 
Including the constant terms we obtain
\ben
2G_1^{(a)}&=& -\frac{1}{N_c}(2\csw^{(0)}-1)L+0.004572(2), \\
2G_1^{(b)}&=& -\frac{N_c}{2}(6\csw^{(0)}-3)L+0.08311(3), \\
2G_1^{(c)}&=&  \frac{N_c}{2}3\csw^{(0)}L-0.08133(3), \\
2G_1^{(d)}&=&  0.29739454(1), \\
2G_1^{(e)}&=&  \frac{1}{2}\left\{-\left(C_F-\frac{1}{2 N_c}\right)
+\left(C_F+\frac{1}{2 N_c}\right)\csw^{(0)}\right\}L-0.017574(1), \\
2G_1^{(f)}&=&  \frac{1}{2}\left\{-\left(C_F-\frac{1}{2 N_c}\right)
+\left(C_F+\frac{1}{2 N_c}\right)\csw^{(0)}\right\}L-0.017574(1), 
\een
where 
\be
L=\frac{1}{16\pi^2}\ln\left|\frac{\pi^2}{\lambda^2 a^2}\right|
\ee
denotes the contribution 
of the infrared divergence with the fictitious gluon mass $\lambda$.
The integrals are numerically estimated by a mode sum for a periodic box of
a size $N^4$ with $N=64$ after transforming the momentum 
variable through $k^\prime_\mu=k_\mu-{\rm sin}k_\mu$.
We choose $\Lambda=\pi$ for the cut-off. 
It is found that the tadpole diagram of Fig.~\ref{fig:vtx_1loop} (d) 
gives the dominant contribution.
The total contribution from infrared divergent terms becomes
\ben
L\times (1-\csw^{(0)})\left\{\frac{3}{2N_c}-C_F+\frac{3N_c}{2}
\right\},
\een
therefore,
the infrared divergences are canceled out in a nontrivial way if and
only if the tree-level coefficient is properly tuned: $\csw^{(0)}=1$.
Whereas the coefficient of the logarithmic infrared divergence
in each diagram is independent of the gauge action, the constant terms 
depend on it. In Table~\ref{tab:c_sw} we present
the results of $\csw^{(1)}$ for the various improved gauge 
actions. 
The value of $\csw^{(1)}$ for DBW2 is obtained for the first time.
Other results are consistent with those obtained by the previous work
employing different infrared regularizations\cite{csw_sf}.

Here we give a brief description on the mean field improvement of $\csw$.
The tadpole contribution of Fig.~\ref{fig:vtx_1loop} is given by
\ben
\csw^\tad&=& g^2 \intlat\left[
\left\{\left(\frac{4}{3}C_F+\frac{2}{3 N_c}\right)
-\left(\frac{3}{2 N_c}-C_F\right)\sin^2\left(\frac{k_\nu}{2}\right)
\right\}D_{\mu\mu}(k)\right.\nn\\
&&\left.-\left(2C_F-\frac{1}{N_c}\right)\sin\left(\frac{k_\mu}{2}\right)
\sin\left(\frac{k_\nu}{2}\right)D_{\mu\nu}(k) \right],
\een
where $\mu$, $\nu$ are unsummed and $\mu\ne\nu$.
The numerical values for the various gauge actions 
are listed in Table~\ref{tab:c_sw}.
The mean field improvement is applied as
\ben
\csw&=&\left(1+\left(\frac{4}{3}C_F+\frac{2}{3 N_c}\right)g^2T_\mf\right)
\left(1+g^2\csw^{(1)}
-\left(\frac{4}{3}C_F+\frac{2}{3 N_c}\right)g^2T_\mf\right)+O(g^4)\nn\\
&\rightarrow&\frac{1}{u^3}\left(1+g^2\csw^{(1)}
-\left(\frac{4}{3}C_F+\frac{2}{3 N_c}\right)g^2T_\mf\right)+O(g^4),
\een
where $u=P^{1/4}$ is evaluated by Monte Carlo simulation.
The derivation of $T_\mf$ 
is given in detail in Sec.~III of Ref.~\cite{dwf_pt_rg}.

The mean-field improved $\msbar$ 
coupling $g_\msbar^2(\mu )$ at the scale $\mu$ is obtained from the lattice
bare coupling $g_0^2$ with the use of the following relation:
\ben
\dfrac{1}{g_{\overline{\rm MS}}^2(\mu )}
&=& \dfrac{P}{g^2_0} + d_g + c_p +\dfrac{22}{16\pi^2} \log (\mu a)
+N_f\left(d_f -\dfrac{4}{48\pi^2} \log (\mu a)\right) .
\label{eq:g2_plaq}
\een
For the improved gauge action 
one may use an alternative formula\cite{cppacs}
\ben
\dfrac{1}{g_{\overline{\rm MS}}^2(\mu )}
&=& \dfrac{c_0 P + 8 c_1 R1+ 16c_2 R2 +8 c_3 R3}{g^2_0} \nn \\
& &+ d_g + (c_0\cdot c_p + 8 c_1\cdot c_{R1}+16 c_2\cdot c_{R2}+
8 c_3\cdot c_{R3})
+\dfrac{22}{16\pi^2} \log (\mu a) \nn\\
&&+N_f\left(d_f -\dfrac{4}{48\pi^2} \log (\mu a)\right),
\label{eq:g2_rg}
\een
where
\ben
P  &=& \frac{1}{3}{\rm Tr} U_{plaquette}    =1 - c_{p} g_0^2 +O(g_0^4),\\ 
R1 &=& \frac{1}{3}{\rm Tr} U_{rectangle}    =1 - c_{R1} g_0^2 +O(g_0^4),\\
R2 &=& \frac{1}{3}{\rm Tr} U_{chair}        =1 - c_{R2} g_0^2 +O(g_0^4),\\
R3 &=& \frac{1}{3}{\rm Tr} U_{parallelogram}=1 - c_{R3} g_0^2 +O(g_0^4),
\een
and the measured values are employed for $P$, $R1$, $R2$ and $R3$. 
The values of $c_{p}$, $c_{R1}$, $c_{R2}$ and $c_{R3}$ for various 
gauge actions are listed in Table~XVI of Ref.~\cite{dwf_pt_rg}.

\begin{figure}[t]
\centering{
\hskip -0.0cm
\includegraphics[width=140mm,angle=0]{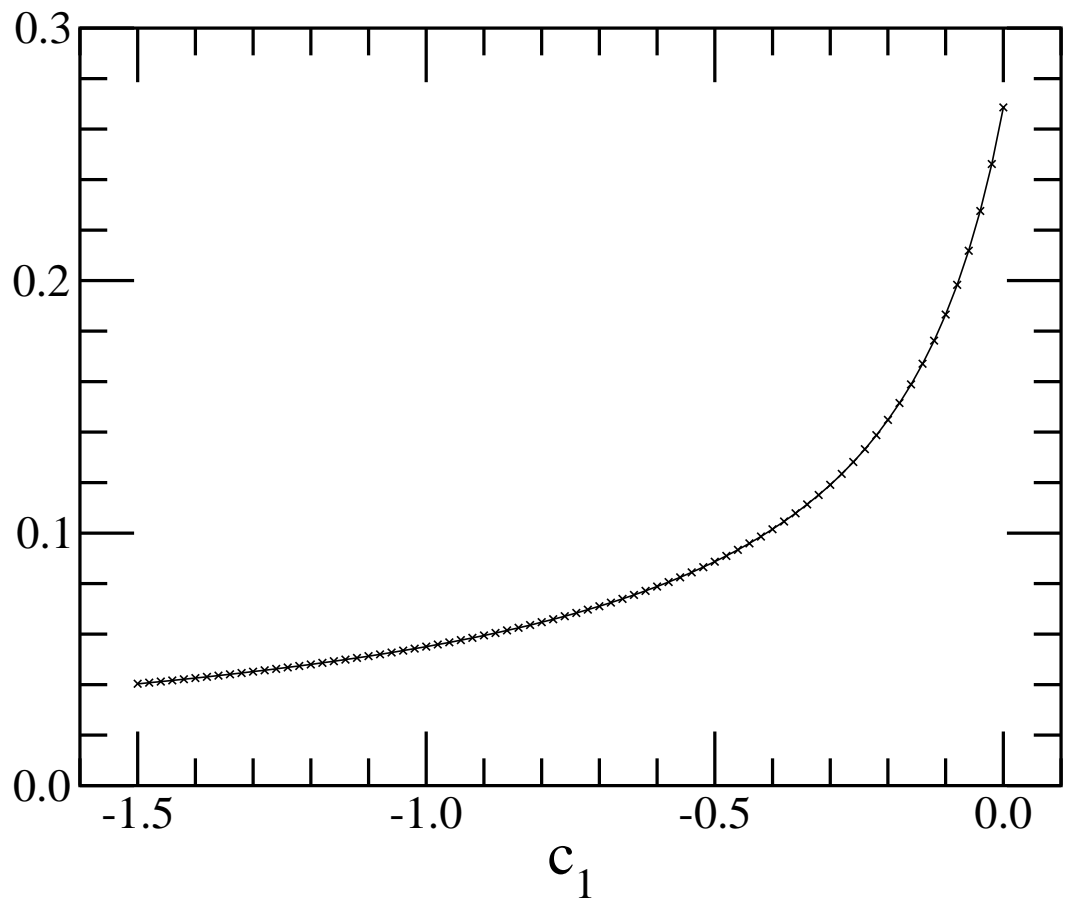}     
}
\caption{$\csw^{(1)}$ as a function of $c_1$ with $c_2=c_3=0$.
Solid line denotes the fitting result of eq.(\protect{\ref{eq:fit_csw}}).}
\label{fig:csw_c1}
\vspace{8mm}
\end{figure}

For later convenience it would be a good idea to
parameterize the value of $\csw^{(1)}$ as a function 
of $c_1$ while keeping $c_2=c_3=0$.
In Fig.~\ref{fig:csw_c1} we plot the results of $\csw^{(1)}$ evaluated
by a mode sum with $N=64$, where $c_1$ is chosen from $-1.5$ to 0
at intervals of 0.02. We observe that $\csw^{(1)}$ seems to be
divergent as $c_1$ increases. 
This behavior is well described by the rational expression,
\ben
\csw^{(1)}&=&\frac{0.26849-0.14193c_1-0.13641c_1^2-0.07996c_1^3-0.01911c_1^4}
{1-5.08365c_1}.
\label{eq:fit_csw}
\een     
where the fitting result is also depicted in Fig.~\ref{fig:csw_c1}.
The difference between the actual value and the fit is less than 0.1\%
for $-1.5\le c_1\le 0$.

\section{Conclusion}

In this paper we determine the clover coefficient $\csw$ 
in the massless SW quark action up to one-loop order 
for the various improved gauge actions employing the conventional 
perturbative method with the fictitious gluon mass as an infrared regulator.
The validity of the method is checked by comparing the results to those
previously obtained by the Schr{\"o}dinger functional method: both show
a good agreement within error bars.
For later convenience
our results are parametrized in terms of the improvement coefficient $c_1$
of the gauge action.
An important application of this conventional perturbative method
is to determine $\csw$ for the massive quarks in
the heavy quark formulation proposed by the authors, where
the relativistic on-shell improvement is extended to the massive case
including any power corrections of $m_Qa$.
Whereas $c_E$ and $c_B$ receive different $m_Q a$ corrections 
in this formulation, 
a modification of the present calculational techniques can be done 
in a straightforward manner\cite{csw_m}.

\section*{Acknowledgments}
S.A. would like to thank Prof. P. Weisz for informative correspondence.
This work is supported in part by the Grants-in-Aid for
Scientific Research from the Ministry of Education, 
Culture, Sports, Science and Technology.
(Nos. 13135204, 14046202, 15204015, 15540251, 15740165).

\newpage

\newcommand{\J}[4]{{ #1} {\bf #2} (#3) #4}
\newcommand{\MPL}{Mod.~Phys.~Lett.}
\newcommand{\IJMP}{Int.~J.~Mod.~Phys.}
\newcommand{\NP}{Nucl.~Phys.}
\newcommand{\PL}{Phys.~Lett.}
\newcommand{\PR}{Phys.~Rev.}
\newcommand{\PRL}{Phys.~Rev.~Lett.}
\newcommand{\AP}{Ann.~Phys.}
\newcommand{\CMP}{Commun.~Math.~Phys.}
\newcommand{\PTP}{Prog.~Theor.~Phys.}
\newcommand{\Suppl}{Prog. Theor. Phys. Suppl.}
\bibliography{basename of .bib file}

\newpage

\begin{table*}[htb]
\caption{One-loop coefficient of $\csw$ for various improved gauge actions. 
Tadpole contribution of 
Fig.~\protect{\ref{fig:vtx_1loop}} (d) is also listed.}
\label{tab:c_sw}
\newcommand{\m}{\hphantom{$-$}}
\newcommand{\cc}[1]{\multicolumn{1}{c}{#1}}
\renewcommand{\tabcolsep}{2pc} 
\renewcommand{\arraystretch}{1.2} 
\begin{tabular}{lllll}
\hline
gauge action   & $c_1$ & $c_3$ & $\csw^{(1)}$ & $\csw^\tad$  \\
\hline
plaquette      & \m0         & \m0      & \m0.26858825(1) & 0.29739454(1)\\
Symanzik       & \m--1/12    & \m0      & \m0.19624449(1) & 0.23543879(1)\\
Iwasaki        & \m--0.331   & \m0      & \m0.11300591(1) & 0.15988461(1)\\
Iwasaki'       & \m--0.27    & \m--0.04 & \m0.12036501(1) & 0.16566349(1)\\
Wilson         & \m--0.252   & \m--0.17 & \m0.10983411(1) & 0.15292225(1)\\
DBW2           & \m--1.40686 & \m0      & \m0.04243181(1) & 0.08997537(1)\\
\hline
\end{tabular}
\end{table*}

\newpage 

\begin{figure}[t]
\begin{center}

\begin{picture}(200,150)(0,0)
\ArrowLine(50,50)(100,100)
\Text(80,65)[l]{$p$}
\ArrowLine(100,100)(50,150)
\Text(80,135)[l]{$q$}
\Vertex(100,100){2}
\Gluon(100,100)(171,100){5}{5}
\Text(196,115)[c]{$k_1$, $\mu$, $A$}
\LongArrow(171,115)(151,115)
\Text(110,30)[c]{(a)}
\end{picture}

\begin{picture}(200,150)(0,0)
\ArrowLine(50,50)(100,100)
\Text(80,65)[l]{$p$}
\ArrowLine(100,100)(50,150)
\Text(80,135)[l]{$q$}
\Vertex(100,100){2}
\Gluon(100,100)(150,150){5}{5}
\Gluon(100,100)(150,50){-5}{5}
\LongArrow(150,135)(136,121)
\Text(175,145)[c]{$k_2$, $\nu$, $B$}
\LongArrow(150,65)(136,79)
\Text(175,60)[c]{$k_1$, $\mu$, $A$}
\Text(110,30)[c]{(b)}
\end{picture}

\begin{picture}(200,150)(0,0)
\ArrowLine(50,50)(100,100)
\Text(80,65)[l]{$p$}
\ArrowLine(100,100)(50,150)
\Text(80,135)[l]{$q$}
\Vertex(100,100){2}
\Gluon(100,100)(150,150){5}{5}
\Gluon(100,100)(171,100){5}{5}
\Gluon(100,100)(150,50){-5}{5}
\LongArrow(150,135)(136,121)
\Text(175,145)[c]{$k_3$, $\tau$, $C$}
\LongArrow(171,115)(151,115)
\Text(196,115)[c]{$k_2$, $\nu$, $B$}
\LongArrow(150,65)(136,79)
\Text(175,60)[c]{$k_1$, $\mu$, $A$}
\Text(110,30)[c]{(c)}
\end{picture}
\vspace{-8mm}

\end{center}
\caption{Momentum assignments for the quark-gluon vertices.}
\label{fig:fr_vtx_qg}
\vspace{8mm}
\end{figure}                                                                   

\newpage 

\newpage

\end{document}